# Maximal Raman enhancement factor (EF) calculations for hot spots at two metallic spheres

Y. BEN-ARYEH

*Technion-Israel Institute of Technology, Physic Department, Israel, Haifa, 32000*

*phr65yb@physics.technion.ac.il*

**The interaction between two metallic spheres with radius R with external electromagnetic (EM) field polarized in the symmetric $z$ direction is described. Solutions of Laplace equation with bi-spherical coordinates are developed. Hot spots are obtained under the condition that the shortest distance between the two spheres surfaces is very small relative to their radius. Boundary conditions are applied which assume very large real negative value for the dielectric constant of the metallic spheres which is valid for atoms like that of $A_u$ or $A_g$ at certain frequencies. Under these conditions the EM field is amplified by many orders of magnitudes relative to the incident EM field. Analytical results for maximal Raman enhancement factor ($EF$) are obtained as function of various parameters. The present study can be applied to surface-enhanced Raman spectroscopy (SERS) and two-photon induced illumination (TPI-PL) in which the amplification is proportional to the fourth power of the incident EM field.**

## 1. INTRODUCTION

An electromagnetic (EM) mechanism for surface enhanced Raman spectroscopy (SERS) involves the localization and amplification of incident light fields by a surface plasmon resonance. Motivated by the need to quantify the EM enhancement within such structures, computational studies using the finite element method (FEM), the finite-difference time domain (FDTD) method, discrete dipole approximation, and the generalized multipole Mie (GMM) analysis appeared as ideal complement to experimental studies [1,2]. The ability of SERS to obtain single-molecule sensitivity relies on the formation of regions with ultra-highly enhancement called hot spots. These highly enhancement sites occur at the junction between two and more plasmonic structures separated by a very small gap.

In the limit of very small particles, the EM interaction between different parts of the metal is instantaneous. Then Maxwell equations are leading to the condition that the electric and magnetic fields are longitudinal $(\vec{\nabla} \times \vec{E} = \vec{\nabla} \times \vec{H} = 0)$. The magnetic response of the particles is negligible at optical frequencies for very small particles so that the electric field is the gradient of the scalar electric potential, $\vec{E} = -\vec{\nabla}\psi$ [3]. In the present work analytical results are developed for maximal SERS enhancement factor ($EF$) under such conditions for two nearby metallic spheres. It is of much importance to find the conditions under which the $EF$ is maximal so that spectroscopic effects on one molecule level can be observed. Although we treat a very special system one can learn from this analysis about the general conditions for getting maximal $EF$.

We consider two metallic spheres of equal radius $R$ described in Fig. 1. We choose the vertical z-axis along the line passing through the centers of the spheres. The perpendicular $x, y$ plane contains the midpoint between the two spheres. We assume that the distance from the center of one sphere with radius, R (the upper one) to the center of the coordinate system along the $z$ coordinate is $+D$ and that for the other sphere with the same radius R (the lower one) is $-D$.

We define

$$D = R + \delta \, ; \, a = \left[D^2 - R^2\right]^{1/2} = \left[(R+\delta)^2 - R^2\right]^{1/2} \quad . \tag{1}$$

The shortest distance between the two spheres surfaces is given by: $2\delta$. For simplicity we treat mainly the case where the incoming EM field is homogenous and the electric field $E_z$ is along the $z$ axis. Assuming certain values for the dielectric constants [4,5] (for the two spheres $\varepsilon(\omega)$ which are function of the frequency $\omega$ and for the surrounding medium $\varepsilon_1$) we present the solutions of the Laplace equation for the limiting case for which $\delta \ll R$. The two focuses $F_1$ and $F_2$ are located at a distance $a$ from the center of the coordinate system along the symmetric $z$ axis, in upper and lower directions, respectively

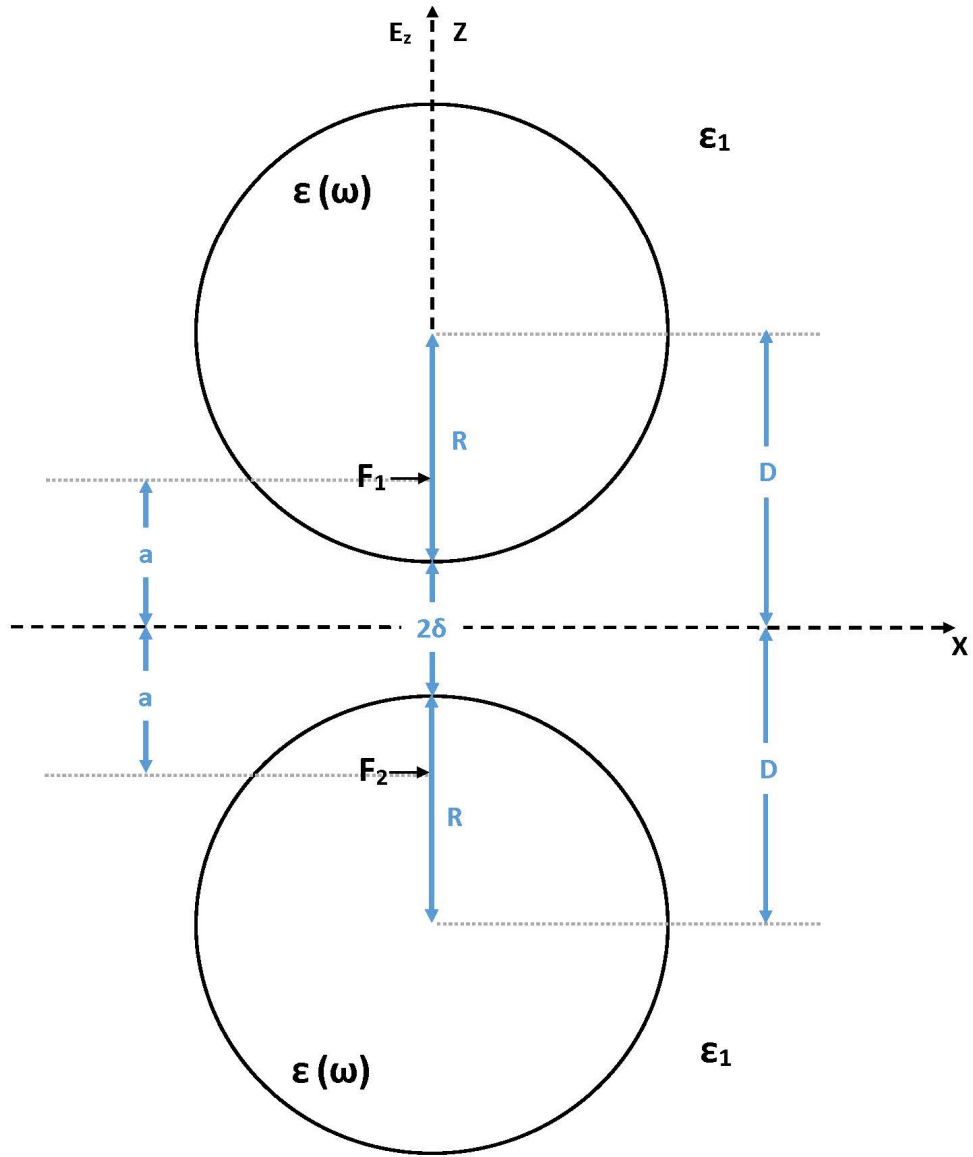

**Fig. 1.** Two spheres with metallic dielectric constant $\varepsilon(\omega)$ with radius R and the surrounding medium with dielectric constant $\varepsilon_1$, under homogenous external EM field $E_z = E_0$. Various parameters are described in the present $x, z$ coordinates system.

Laplace equations solutions for the upper and lower spheres are given by $\psi_+$ and $\psi_-$ ,respectively, and Laplace equation solution for the surrounding medium is given by $\psi_1$. The present system has a cylindrical symmetry under rotation around the $z$ axis. Thus, the two focuses are not changed by this rotation. It was shown [6,7] that Raman signals are strongly amplified when the molecules are inserted in the interstitial gaps between nanoparticles due to the very strong EM fields induced in these gaps ("hot spots"). Various experimental results on surface enhanced Raman spectroscopy (SERS) from molecules on aggregates of nanoparticles are interpreted by hot spot mechanism [8,9]. Special studies were made on Raman signals enhancement in dimers (two nanoparticles) [10.11]. It was found that the Raman signals of spherical dimers are strongly enhanced when the incident polarization is parallel to the inter particle axis of the dimer (parallel polarization) [12]. In this case the opposite charges of polarization are facing each other at the small gap and by their interaction generate a huge EM field. On the other hand, when the incident EM field is polarized in direction perpendicular to the inter particle axis (perpendicular polarization) the induced charges are in directions different from that of the gap. In this case, individual local surface plasmons (LSP) in the dimer do not interact strongly with each other. As a result, EM field interaction is approximately compared in this case with that of isolated particles. It was found that the signal in SERS is proportional to the fourth power of the amplified EM field for parallel polarization. Similar results are obtained by two-photons-induced luminescence (TPI-PL) [13]. Raman scattering and TPI-PL phenomena are increased by many orders of magnitude relative to that of the ordinary ones, for molecules inserted in these hot spots.

Laplace equation solutions for single metallic sphere interacting with homogenous EM field leads to electrostatic field of a dipole located at the center of the sphere (e.g. [14]). We treat in the present analysis the case of parallel polarization in dimers in which the induced EM strength depends strongly on the inter particle distance. For cases in which the distance been the two spheres is relatively large i.e., when $2\delta \geq R$ the interaction between the two spheres can be treated by conventional theories about dipole-dipole interactions [15]. Experimental results are in agreement with such theory (e.g. [16]). It was found also that fabricated nano-shells can provide SERS enhancement compared to dimers [17]. Various structures and materials for SERS field enhancement were described [18-20].

For cases in which the distance between the two spheres is very short i.e., when $2\delta \ll R$, different plasma resonances become important, and the analysis by Laplace equation for such dimers becomes quite complicated [21-27]. In the present work we study the solutions of Laplace equation solutions for dimers with bi-spherical coordinates [23-27] which are developed under the condition $2\delta \ll R$. Hot spots are produced in the system of two metallic spheres interacting with external homogeneous EM field. While important results (mainly for the potential) for the present system were developed by solving Laplace equation with the use of bi-spherical coordinates the analysis for the hot spots remained problematic due to convergence problems. By using boundary conditions various authors [23-26] obtained after some tedious algebra set of recursion relations (or equivalently infinite set of linear equations) for the Laplace equation superposition solutions. Such system was truncated by taking finite set of linear equations and was solved on computers. Special care was taken to make sure its convergence i.e., that the number of recursion relations is not too small (especially for nearby spheres where very high number of recursion relations is needed). We give here an alternative for deriving the EM fields at the hot spots by using bi-spherical coordinates with certain approximations. We develop in the present work a relatively simpler model for analyzing the properties of the EM fields by using approximations which are suitable for treating the hot spots with the use of bi-spherical coordinates. Analytical results for maximal enhancement factor ($EF$) are developed.

## 2. DEFINITIONS AND PROPERTIES OF BI-SPHERICAL COORDINATES [22-27]

The bi-spherical coordinates are a special three-dimensional orthogonal coordinates system defined by coordinates $\eta, \alpha, \phi$ [22-26]

$$\begin{aligned}
x &= a\sin\alpha\cos\phi/(\cosh\eta - \cos\alpha) \ , \\
y &= a\sin\alpha\sin\phi/(\cosh\eta - \cos\alpha) \ , \\
z &= a\sinh\eta/(\cosh\eta - \cos\alpha) \\
r &= \sqrt{x^2 + y^2 + z^2} = a\sqrt{\frac{\cosh\eta + \cos\alpha}{\cosh\eta - \cos\alpha}}
\end{aligned} \qquad (2)$$

The inverse transformations of Eq. (2) are given by

$$\sinh \eta = \frac{2az}{\sqrt{\left(x^2 + y^2 + z^2 + a^2\right)^2 - (2az)^2}} \quad , \quad \tanh \eta = \frac{2az}{x^2 + y^2 + z^2 + a^2}$$

$$\cos \alpha = \frac{x^2 + y^2 + z^2 - a^2}{\sqrt{\left(x^2 + y^2 + z^2 + a^2\right)^2 - (2az)^2}} \quad , \quad \tan \alpha = \frac{2a\left(x^2 + y^2\right)^{1/2}}{x^2 + y^2 + z^2 - a^2} \quad . \tag{3}$$

$$\tan \phi = y / x$$

The two poles with $\eta = \pm\infty$ are located on the $z$ axis at $z = \pm a$ and denoted in Fig. 1 by $F_1$ and $F_2$. Surfaces of constant $\eta$ are given by the spheres

$$x^2 + y^2 + (z - a \coth \eta)^2 = \frac{a^2}{\sinh^2 \eta} \quad . \tag{4}$$

For constant value of $\eta$, Eq. (4) describes spheres. The special value $\eta = +\eta_0$ is defined by the following equivalent equations

$$\sinh \eta_0 = a / R \; ; \; \cosh \eta_0 = \frac{1}{R}\sqrt{R^2 + a^2} \; ; \; D = a \coth \eta_0 \quad . \tag{5}$$

By substituting, in Eq. (4) the special value $\eta = \eta_0$, from Eq. (5), we get:

$$x^2 + y^2 + (z - D)^2 = R^2 \quad . \tag{6}$$

This equation for $\eta = \eta_0$ represents in Fig. 1 the upper sphere with radius $R$ where its center is moved from the center of the coordinate system by a distance D in the positive $z$ direction. For the special case with $\eta = -\eta_0$ we can use Eqs. (4-5) with $\sinh \eta_0 \to \sinh(-\eta_0) = -\sinh \eta_0$. Then instead of Eq. (6) we get

$$x^2 + y^2 + (z + D)^2 = R^2 \tag{7}$$

where this equation represents the lower sphere for $\eta = -\eta_0$ with radius $R$ where its center is moved from the center of the coordinate system by a distance D in the negative $z$ direction.

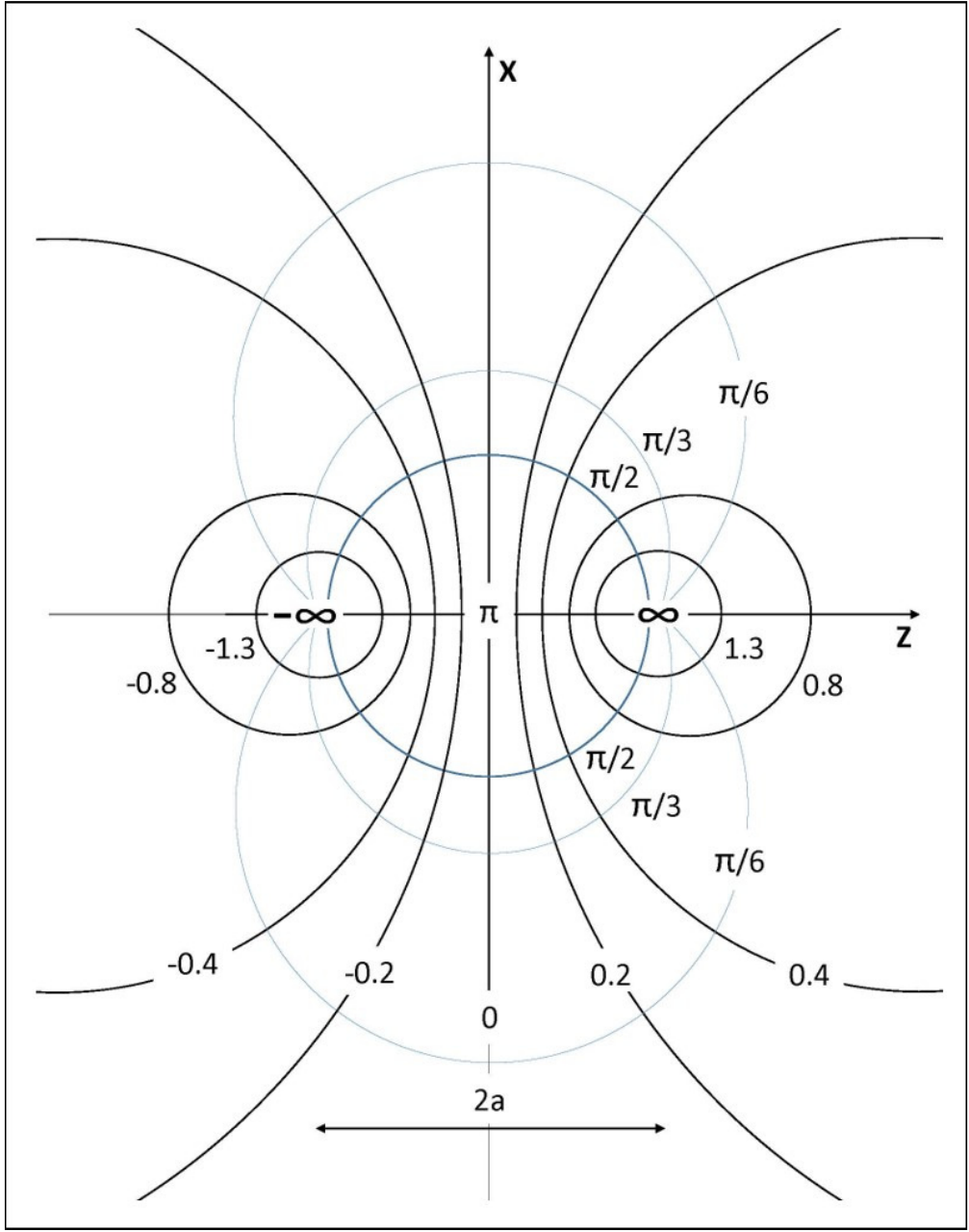

**Fig. 2.** Bi-spherical coordinates in the $x, z$ plane showing circles of constant bi-spherical radial coordinate $\eta$ where curves of constant polar angular $\alpha$ with $\alpha = \pi, \pi/2, \pi/3, \pi/6$ are perpendicular to these curves [3]. EM field polarized in the $z$ direction propagates in the $x$ direction. The curve for $\alpha = \pi$ is along the $z$ axis.

We treat in the present article the special case for which the two spheres have equal radius and when the incident uniform EM field is parallel to the $z$ direction. In this case the electrostatic potential has cylindrical symmetry about the $z$ axis. It is therefore independent of the angle $\phi$ and only the term $m = 0$ is retained [22-26]. In Fig. 2 we describe the coordinates $\eta$ for certain values of $\eta$ as function of the $x, z$ coordinates. The large circles represented by small values of $\eta$ are truncated in this figure. For special values $+\eta_0$ and $-\eta_0$ the circles are given by Eqs. (6-7). By their rotation around the symmetric $z$ axis they describe the two spheres of the present system. The conditions for $\eta_0$ are fixed by Eq. (5). For complete description of the bi-spherical coordinates we need to add the $\alpha$ coordinates which are orthogonal to the $\eta$ coordinates.

## 3. LAPLACE-EQUATION SOLUTIONS FOR TWO METALLIC SPHERES WITH INCIDENT EM FIELD PARALLEL TO THE SYMMETRIC Z COORDINATE

We define $\psi_+, \psi_-$ and $\psi_1$, respectively, as the potentials (with the condition $m = 0$) inside the upper sphere, the lower sphere, and the surrounding medium, respectively. The potential due to the external field $V_0$ is assumed to be given by $V_0 = -E_0 z$. In the present article the external field . is written in short notation as $E_0$. It is antisymmetric with respect to reflections through the $xy$ plane: $z \to -z$ or $\eta \to -\eta$.

The potential $\psi_1$ outside the spheres is given with the same symmetry as that of the external field potential [23,27]:

$$\psi_1(\eta,\alpha) = (\cosh\eta - \cos\alpha)^{1/2} \sum_{n=0}^{\infty} A_n \sinh\left[\left(n+\frac{1}{2}\right)\eta\right] P_n(\cos\alpha) - \frac{E_0 a \sinh\eta}{\cosh\eta - \cos\alpha} \quad ;$$

$$\frac{a \sinh\eta}{\cosh\eta - \cos\alpha} = z$$

(8)

By using the relation [23,27]

$$z = \pm a (\cosh\eta - \cos\alpha)^{1/2} \sum_{n=0}^{\infty} \sqrt{2}(2n+1) P_n(\cos\alpha) e^{\mp\eta(n+1/2)} \qquad (9)$$

where the upper signs hold for positive $z$, and the lower signs hold for negative $z$, Eq. (8) is transformed to

$$\psi_1(\eta,\alpha) = (\cosh\eta - \cos\alpha)^{1/2} \sum_{n=0}^{\infty} P_n(\cos\alpha) \left[ A_n \sinh\left[\left(n+\frac{1}{2}\right)\eta\right] - E_0 a 2^{1/2}(2n+1) e^{-\eta(n+1/2)} \right] \cdot \qquad (10)$$

For $\eta = \eta_0$, and positive $z$, Eq. (10) is transformed to

$$\psi_1(\eta_0,\alpha) = (\cosh\eta_0 - \cos\alpha)^{1/2} \sum_{n=0}^{\infty} P_n(\cos\alpha) \left[ A_n \sinh\left[\left(n+\frac{1}{2}\right)\eta_0\right] - E_0 a 2^{1/2}(2n+1) e^{-\eta_0(n+1/2)} \right] \cdot \qquad (11)$$

From the fact that $\psi_+$ and $\psi_-$ have to be finite at the points: $x = y = 0; z = \pm a$, where: $\eta = \pm\infty$, we obtain [23,27]

$$\psi_+(\eta,\alpha) = (\cosh\eta - \cos\alpha)^{1/2} \sum_{n=0}^{\infty} B_n \exp^{-\left(n+\frac{1}{2}\right)\eta} P_n(\cos(\alpha)) \quad , \qquad (12)$$

$$\psi_-(\eta,\alpha) = -(\cosh\eta - \cos\alpha)^{1/2} \sum_{n=0}^{\infty} B_n \exp^{\left(n+\frac{1}{2}\right)\eta} P_n(\cos(\alpha)) \quad . \quad . \qquad (13)$$

One should notice that the function $\sinh\left[\left(n+\frac{1}{2}\right)\eta\right]$ is antisymmetric with respect to reflections through the $xy$ plane i. e. $z \to -z$ or $\eta \to -\eta$ in agreement with the symmetry of the external electric field. The general solutions for the potentials in the surrounding medium, and in the upper sphere are given by Eqs. (10) and (12), respectively. But the coefficients $A_n$ and $B_n$ should be obtained from the boundary conditions.

Using Eqs. (10-13) we get the EM potentials as function of the bi-spherical coordinates. Transformation of these equations to be functions of the $x, y, z$ coordinates can be made by using Eqs. (2-3). The coefficients $A_n$ and $B_n$ are treated as follows:

We use the first boundary condition given as

$$\psi_+(\eta_0) = \psi_1(\eta_0) \quad . \tag{14}$$

By using the equality (14) and comparing the corresponding expressions (11) and (12) for $\eta = \eta_0$ we obtain

$$B_n \exp^{-\left(n+\frac{1}{2}\right)\eta_0} = A_n \sinh\left[\left(n+\frac{1}{2}\right)\eta_0\right] - 2^{1/2} E_0 a (2n+1) e^{-(n+1/2)\eta_0} \quad . \tag{15}$$

A second boundary condition can be used as [23-27]:

$$\varepsilon(\omega)\left[\frac{\partial \psi_+(\eta,\alpha)}{\partial \eta}\right]_{\eta=\eta_0} = \varepsilon_1 \left[\frac{\partial \psi_1(\eta,\alpha)}{\partial \eta}\right]_{\eta=\eta_0} \quad . \tag{16}$$

Using Eq. (10) we get

$$\varepsilon_1 \left\{\frac{\partial}{\partial \eta}\psi_1(\eta,\alpha)\right\}_{\eta=\eta_0} =$$
$$\varepsilon_1 \left[\frac{\partial}{\partial \eta}\left\{(\cosh\eta - \cos\alpha)^{1/2} \sum_{n=0}^{\infty} P_n(\cos\alpha)\left[A_n \sinh\left[\left(n+\frac{1}{2}\right)\eta\right] - E_0 a 2^{1/2}(2n+1)e^{-\eta(n+1/2)}\right]\right\}\right]_{\eta=\eta_0} \tag{17}$$

Using Eq. (12) we get:

$$\varepsilon(\omega)\left[\frac{\partial \psi_+(\eta,\alpha)}{\partial \eta}\right]_{\eta=\eta_0}$$
$$= \varepsilon(\omega)\left[\frac{\partial}{\partial \eta}\left\{(\cosh\eta - \cos\alpha)^{1/2} \sum_{n=0}^{\infty} B_n \exp^{-\left(n+\frac{1}{2}\right)\eta} P_n(\cos\alpha)\right\}\right]_{\eta=\eta_0} \quad . \tag{18}$$

The derivatives $\left[\frac{\partial \psi_+(\eta,\alpha)}{\partial \eta}\right]_{\eta=\eta_0}$ and $\left\{\frac{\partial}{\partial \eta}\psi_1(\eta,\alpha)\right\}_{\eta=\eta_0}$ include the local plasmons charges induced on the surfaces of the spheres. In the present work we follow the idea, that for treating the limits of large field enhancement in hot spots we can use the following approximation which will simplify very much the analysis:

1) Derivatives in Eqs. (17) and (18) include derivatives according to $\eta$ of $(\cosh\eta - \cos\alpha)^{1/2}$ in addition to the derivatives of the terms in the summations of these equations. Under the condition that $\delta$ is much smaller than $R$ there are many $B_n$ and $A_n$ terms including exponential terms with derivatives proportional to the integer $n$ which are very large relative to the derivatives of $(\cosh\eta - \cos\alpha)^{1/2}$ so that the latter derivatives can be neglected for hot spots. The terms with larger value of $n$ for $\left[\dfrac{\partial \psi_+(\eta,\alpha)}{\partial \eta}\right]_{\eta=\eta_0}$ and $\left\{\dfrac{\partial}{\partial \eta}\psi_1(\eta,\alpha)\right\}_{\eta=\eta_0}$ represent more rapid decay of the local surface plasmons. 2) In the present analysis for hot spots we assume that $A_n$ and $B_n$ are very large numbers so that for the purpose of using the boundary condition (16) the small term $E_0 a 2^{1/2}(2n+1)e^{-\eta(n+1/2)}$ can be neglected.

By substituting Eqs. (18) and (17) into Eq. (16) and using the above approximations we get for the relation between $A_n$ and $B_n$:

$$\varepsilon_1 A_n \left[\cosh\left(n+\frac{1}{2}\right)\eta_0\right] = -\varepsilon(\omega) B_n \exp\left[-\left(n+\frac{1}{2}\right)\eta_0\right] \quad ; \tag{19}$$

Here the common factors: $(\cosh\eta - \cos\alpha)^{1/2}$, $P_n(\cos\alpha)$, $\left(n+\dfrac{1}{2}\right)$ and the term proportional to $E_0$ were neglected. Eq. (19) shows that for larger values of $-\varepsilon(\omega)$ the term $B_n$ becomes smaller. One should notice that while Eq. (19) represents an approximate relation for hot spots, Eq. (15) is exact one.

By substituting Eq. (19) for $B_n \exp\left[-\left(n+\dfrac{1}{2}\right)\eta_0\right]$ into Eq. (15) we get:

$$\dfrac{\varepsilon_1}{-\varepsilon(\omega)}\left\{A_n\left[\cosh\left(n+\frac{1}{2}\right)\eta_0\right]\right\}$$
$$= \left\{A_n \sinh\left[\left(n+\frac{1}{2}\right)\eta_0\right] - E_0 a 2^{1/2}(2n+1)e^{-\eta(n+1/2)}\right\} \tag{20}$$

By rearranging the terms in Eq. (20) we get:

$$A_n = \frac{2^{1/2}(2n+1)\varepsilon(\omega)E_0 a e^{-(n+1/2)\eta_0}}{\left\{\varepsilon_1 \cosh\left(n+\frac{1}{2}\right)\eta_0 + \varepsilon(\omega)\sinh\left(n+\frac{1}{2}\right)\eta_0\right\}} \; ; \; for \delta \ll R \tag{21}$$

By transforming the hyperbolic functions of Eq. (21) to exponential terms we get:

$$A_n = \frac{2^{1/2}(2n+1)\varepsilon(\omega)E_0 a e^{-(n+1/2)\eta_0}}{\left\{\left[\frac{\varepsilon_1+\varepsilon(\omega)}{2}\right]e^{(n+1/2)\eta_0} + \left[\frac{\varepsilon_1-\varepsilon(\omega)}{2}\right]e^{-(n+1/2)\eta_0}\right\}} \; ; \; for \delta \ll R \; .$$

We divide both numerator and denominator of this equation by $e^{-(n+1/2)\eta_0}$. Then we get

$$A_n = \frac{2^{3/2}(2n+1)\varepsilon(\omega)E_0 a}{\left\{[\varepsilon_1+\varepsilon(\omega)]e^{(2n+1)\eta_0} + [\varepsilon_1-\varepsilon(\omega)]\right\}} \; ; \; for \delta \ll R \; . \tag{22}$$

We note that the calculation of the coefficients $A_n$ by the use of Eq. (22) becomes quite simple as it can be derived in a straight forward way by the use of the parameter $\eta_0$ and the experimental parameters: $\varepsilon(\omega)$, and $\varepsilon_1$. The use of the present approach is limited, however, by the validity of Eq, (19) appropriate to hot spots. For more accurate calculations we should add the contribution of the derivatives of $(\cosh\eta-\cos\alpha)^{1/2}$ but this will complicate very much the analysis without any analytical results for EM field enhancement (EF) (see e.g. [23-27]). For getting maximal enhancement factor ($EF$) one uses metals of the type of $Au$ or $Ag$ which at certain frequencies $\varepsilon(\omega)$ is very large real negative value (taken as experimental parameter) with negligible imaginary value. The present calculations would lead therefore in the following Sections to the upper limit of enhancement factor ($EF$) [28-37] under the above conditions and approximations.

## 4. THE EM FIELD IN BI-SPHERICAL COORDINATES AT THE HOT SPOTS AND IT TRANSFORMATION TO $x,z$ COORDINATES

Solutions, of Laplace equation are simplified by using the bi-spherical coordinates in the $x,z$ plane (in Fig.1 where $y=0$). The bi-spherical coordinates result from rotating this two-dimensional coordinate system about the symmetric $z$ axis that connects the two focuses. Thus, the two focuses in bipolar coordinates remain points on the $z$ axis, (the axis of rotation) in the bi-spherical coordinate system. This electrostatic potential has cylindrical symmetry about the $z$ axis, so that it is independent of this rotation (for $m=0$). Then we can transform the solutions, which are function of bi-spherical coordinates to that of the $x,z$ coordinates assuming $y=0$.

The normal component of the EM field $\vec{E}$ for which $m=0$ [23] is related in the space outside of the two spheres to the gradient in bi-spherical coordinates given as:

$$-\vec{E} = grad\ \psi_1 = \frac{\cosh\eta - \cos\alpha}{a}\left(\hat{a}_\eta \frac{\partial}{\partial \eta} + \hat{a}_\alpha \frac{\partial}{\partial \alpha}\right)\psi_1 \quad . \tag{23}$$

where $\hat{a}_\eta$, $\hat{a}_\alpha$ are unit vectors in the $\eta, \alpha$ directions, respectively, i.e. in the bi-spherical radial direction $\hat{a}_\eta$ and in direction perpendicular to $\hat{a}_\eta$

Since the potential $\psi_1$ in bi-spherical coordinates is given in Eq. (10) by sum of $n$ terms, the gradient in the normal $\eta$ direction is given by

$$-E_\eta = \left[(\cosh\eta - \cos\alpha)/a\right]\frac{d}{d\eta}\sum_n \psi_n \quad , \quad \sum_n \psi_n = \psi_1 \quad . \tag{24}$$

In the derivation for gradient of the potential for the normal component (in the radial direction) only derivatives relative to $\eta$ are taken into account while $\eta_0$ and $\alpha$ remain certain constants. By operating with Eq, (24) on $\psi_1(\eta,\alpha)$ of Eq. (10) we take into account only the derivatives of the terms proportional to $A_n$ representing the amplified potential which is very large relative to the external potential terms $-E_0 z$. Then we get:

$$-E_\eta = \left[(\cosh\eta - \cos\alpha)/a\right]\frac{d}{d\eta}\left[(\cosh\eta - \cos\alpha)^{1/2} \sum_{n=0}^{\infty} A_n \sinh\left[\left(n+\frac{1}{2}\right)\eta\right]P_n(\cos\alpha)\right] \quad . \tag{25}$$

Since the derivative of $(\cosh\eta - \cos\alpha)^{1/2}$ relative to $\eta$ is very small relative to the derivatives of the sinh functions (for $\delta \ll R$) where the number of coefficients $A_n$ is very large we neglect this derivative and get

$$-E_\eta = \left[(\cosh\eta - \cos\alpha)^{3/2}/a\right]\left[\sum_{n=0}^{\infty}(n+\frac{1}{2})\left[\cosh\left(n+\frac{1}{2}\right)\eta\right]A_n P_n(\cos\alpha)\right] \quad . \tag{26}$$

Eq. (26) gives the general solution for the radial EM field in bi-spherical coordinates for hot spots for which $\delta \ll 1$ and for which the coefficient $A_n$ are given by Eq. (21) or (22).

As the hot spots are produced in dimers on (or near) the symmetric axis for which $x = y = 0$ we obtain by using Eqs. (3) for the symmetric $z$ axis:

$$\sin\alpha = 0 \; ; \; \tanh\eta = \frac{\sinh\eta}{\sqrt{1+\sinh^2\eta}} = \frac{2az}{z^2+a^2} \; ;$$

$$\sinh\eta = \frac{2az}{a^2-z^2} \; ; \; \cosh\eta = \frac{a^2+z^2}{a^2-z^2} \; ; \; e^{\pm\eta} = \cosh\eta \pm \sinh\eta = \frac{(a\pm z)^2}{a^2-z^2} = \frac{a\pm z}{a\mp z} \quad (27)$$

At the focal points, for which: $z = \pm a$, the functions $\sinh\eta, \cosh\eta$ and $e^\eta$, are diverging to $\infty$. Substituting the values of $\sinh\eta$ and $\cosh\eta$ from Eq. (27) into $z$ of Eq. (2) we get

$$z = \frac{a\sinh\eta}{\cosh\eta - \cos\alpha} = \frac{2a^2 z}{a^2+z^2-(a^2-z^2)\cos\alpha} \rightarrow \cos\alpha = -1 \quad . \quad (28)$$

This result is demonstrated also in Figure 2 where the curve for $\alpha = \pi$ coincides with the $z$ coordinate in the gap between the metallic surfaces. So, the curve $\alpha = \pi$ leads to the approximation $\cos\alpha = -1$ representing the central part of hot spots. This result leads to special values of the Legendre polynomials on the symmetric $z$ axis given by

$$P_n(\cos(\alpha)) = P_n(-1) = (-1)^n = \begin{cases} 1 & \text{for any even number } n \\ -1 & \text{for any odd number } n \end{cases} \quad . \quad (29)$$

By substituting the value $\cos\alpha = -1$ and Eq. (29) into Eq. (26) we obtain the result for the EM field in bi-spherical coordinates on the symmetric coordinate $z$ including the hot spot:

$$E_{spot} = \sum_{n=0}^{\infty} E_n$$

$$E_n = -\left[(\cosh\eta+1)^{3/2}/a\right]\left[(n+\frac{1}{2})\cosh\left[\left(n+\frac{1}{2}\right)\eta\right]A_n\{(-1)^n\}\right] \quad (30)$$

We are interested in calculations of the total EM field intensity at the hot spot given by $E_{spot}^2$. We notice that in the calculation of $E_{spot}^2$ we have non-diagonal products $E_n E_{n'} (n \neq n')$ with alternating signs so that their total contribution approximately vanishes. We consider therefore only the diagonal incoherent elements. Then for the electric field amplified factor $E_{spot}^{\ 2}$ and for the SERS measurements which are proportional to $E_{spot}^{\ 4}$ we get:

$$E_{spot}^{\ 2} = \left[\sum_{n=0}^{n_{max}} |E_n|^2 \right] \ ;$$

$$|E_n| = \left[(\cosh\eta + 1)^{3/2} / a\right]\left[(n+\frac{1}{2})\cosh\left[\left(n+\frac{1}{2}\right)\eta\right]A_n\right] \ ; \qquad (31)$$

$$E_{spot}^{\ 4} = \left[\sum_{n=0}^{n_{max}} |E_n|^2 \right]^2$$

We should take into account that $E_{spot}^{\ 2}$ gives the electric field squared at the hot spot where products of $E_n$ with $E_{n'}$ ($n \neq n'$) vanish due to the approximation made after Eq. (30). We should take into account also that SERS measurements depend on $E_{spot}^{\ 4}$ so it is obtained by the square of the sum of Eq. (31) (as demonstrated later in the numerical calculations and discussed in Section 5).

We inserted in Eqs. (31) the maximal value $n_{max}$ which guarantees summation convergence but in the numerical calculations we allow this value to tend to $\infty$. We transform Eqs. (31) as a function of the $z$ coordinate by using the following relation from Eq. (27):

$$\cosh(\eta) = \frac{1}{2}\left[\frac{(a+z)}{a-z} + \frac{a-z}{(a+z)}\right] \qquad . \qquad (32)$$

Using this relation, in Eqs. (31) we get

$$|E_n| = \left(\frac{1}{a}\right)\left\{\frac{1}{2}\left[\frac{(a+z)}{a-z} + \frac{a-z}{(a+z)}\right] + 1\right\}^{3/2}$$

$$\left[\left(n+\frac{1}{n}\right)\left(\frac{1}{2}\right)\left\{\left[\frac{a+z}{a-z}\right]^{n+1/2} + \left[\frac{a-z}{a+z}\right]^{(n+1/2)}\right\}A_n\right] \qquad . \qquad (33)$$

We obtain the high EM fields at the hot spots by substituting Eq. (22), in Eq. (33). as demonstrated by numerical calculations in the next section. The number of coefficients $A_n$ needed in the present analysis increases for lower values of $\eta_0$ (corresponding to lower values of $\delta$) but their calculation by using Eq. (22) is quite simple in comparison to the complicated calculations of these coefficient made by truncation of very large number of linear equations used by other authors [23-27]. In the center of the coordinate system Eq. (33) is reduced to simpler form given by

$$|E_n| = \left(\frac{1}{a}\right) 2^{3/2} (n + \frac{1}{2}) A_n \qquad . \qquad (34)$$

## 5. ANALYTICAL RESULTS FOR MAXIMAL FIELD-ENHANCEMENT ($EF$) AT THE CENER OF THE HOT SPOT

The electric field $|E_n|$ at the center of the hot spot is obtained by inserting Eq. (22) into Eq. (34) with summation over $n$. Then we get:

$$|E_n| = 2^{3/2} \sum_{n=0}^{\infty} (n + \frac{1}{2}) \frac{2^{3/2}(2n+1)\varepsilon(\omega) E_0}{\left\{[\varepsilon_1 + \varepsilon(\omega)] e^{(2n+1)\eta_0} + [\varepsilon_1 - \varepsilon(\omega)]\right\}} \qquad . \qquad (35)$$

We define the light intensity as $I = \left[\sum_{n=0}^{\infty} |E_n|^2\right]$. Then the amplification of the light intensity at the center of the hot spots is given by:

$$\left(\frac{I}{|E_0|^2}\right) = \left[\sum_{n=0}^{\infty} \left|\frac{E_n}{E_0}\right|^2\right] = \left\{2^{3/2} \sum_{n=0}^{\infty} (n + \frac{1}{2}) \frac{2^{3/2}(2n+1)\varepsilon(\omega)}{\left\{[\varepsilon_1 + \varepsilon(\omega)] e^{(2n+1)\eta_0} + [\varepsilon_1 - \varepsilon(\omega)]\right\}}\right\}^2 =$$

$$2^6 \sum_{n=0}^{\infty} (n + \frac{1}{2})^2 \frac{(2n+1)^2 \left[\varepsilon(\omega)/(\varepsilon_1 + \varepsilon(\omega))\right]^2}{\left(e^{(2n+1)\eta_0} + G(\omega)\right)^2} \quad ; \quad G(\omega) = [\varepsilon_1 - \varepsilon(\omega)]/[\varepsilon_1 + \varepsilon(\omega)]. \qquad (36)$$

Eq. (36) can be converted approximately to the following integral:

$$\frac{I}{|E_0|^2} = \int_{n=0}^{n=\infty} \left|\frac{E_n}{E_0}\right|^2 dn = \left[\varepsilon(\omega)/(\varepsilon_1 + \varepsilon(\omega))\right]^2 2^{10} \int_0^{\infty} \frac{(n+1/2)^4}{\left(e^{(n+1/2)2\eta_0} + G(\omega)\right)^2} dn . \qquad (37)$$

Eq. (37) is transformed by using the definitions:

$$(n+1/2)2\eta_0 = x \quad ; \quad dx = 2\eta_0 dn \quad . \tag{38}$$

and given approximately as

$$\frac{I}{|E_0|^2} = \int_{n=0}^{n=\infty} \left|\frac{E_n}{E_0}\right|^2 dn = \left[\varepsilon(\omega)/(\varepsilon_1+\varepsilon(\omega))\right]^2 \frac{2^{10}}{(2\eta_0)^5} \int_0^{\infty} \frac{x^4}{\left(e^x + G(\omega)\right)^2} dx \quad . \tag{39}$$

By assuming a very large real negative value of $\varepsilon(\omega)$ we get from Eq. (36) the approximation $G(\omega) \to -1$, Then by using this approximation in Eq. (39) we get:

$$\frac{I}{|E_0|^2} = \left[\varepsilon(\omega)/(\varepsilon_1+\varepsilon(\omega))\right]^2 \frac{2^{10}}{(2\eta_0)^5} \int_0^{\infty} \frac{x^4}{\left(e^x - 1\right)^2} dx =$$

$$\left[\varepsilon(\omega)/(\varepsilon_1+\varepsilon(\omega))\right]^2 \frac{2^{10}}{(2\eta_0)^5} \Gamma(5)\left[\varsigma(4) - \varsigma(5)\right] \tag{40}$$

The integral in Eq. (40) is obtained by using the corresponding integral from Gradshtein and Ryzhik book [38] where $\Gamma(n)$ is the Gamma Function and $\varsigma(n)$ is the Riemann Zeta Function with the values

$$\Gamma(5) = 24 \quad ; \quad \varsigma(4) = 1.08232323 \quad ; \quad \varsigma(5) = 1.03692775 \quad . \tag{41}$$

Inserting these values in Eq. (40) we get:

$$\frac{I}{|E_0|^2} = \int_{n=0}^{n=\infty} \left|\frac{E_n}{E_0}\right|^2 dn = \left[\varepsilon(\omega)/(\varepsilon_1+\varepsilon(\omega))\right]^2 \frac{2^5 \cdot 24 \cdot 0.0454}{\eta_0^5} = \left[\varepsilon(\omega)/(\varepsilon_1+\varepsilon(\omega))\right]^2 \frac{34.87}{\eta_0^5} \quad . \tag{42}$$

As the amplified field in SERS measurements is proportional to the fourth power its field enhancement factor ($EF$) is given by the square of Eq. (42) i.e.,

$$EF = \left[\varepsilon(\omega)/(\varepsilon_1+\varepsilon(\omega))\right]^4 \left(\frac{34.87}{\eta_0^5}\right)^2 = \left[\varepsilon(\omega)/(\varepsilon_1+\varepsilon(\omega))\right]^4 \left(\frac{1216}{\eta_0^{10}}\right) \quad . \tag{43}$$

Eqs. (42-43) represent very fundamental results by which the maximal enhanced light amplification factor for symmetric metallic dimers is proportional to $\eta_0^{-5}$ and the $EF$ for SERS measurements is proportional to $\eta_0^{-10}$. These analytical results are valid under the conditions $2\delta \ll R$, and $G(\omega) = [\varepsilon_1 - \varepsilon(\omega)]/[\varepsilon_1 + \varepsilon(\omega)]) \to -1$. We would like to emphasize that in the present model the denominator $(e^x - 1)^2$ in Eq. (40) leads to resonance condition for certain $n$ value for which $e^{(2n+1)\eta_0} - 1 \to 0$. For cases in which $G(\omega) = \left|[\varepsilon_1 - \varepsilon(\omega)]/[\varepsilon_1 + \varepsilon(\omega)]\right| > 1$ the resonance condition is changed reducing the integral very much as verified by numerical calculation. An important conclusion from the present analysis is that for getting maximal *EF* we need to use metals which have large real negative value for $\varepsilon(\omega)$ i.e. using metals like $A_u$ or $A_g$ at certain frequencies.

For hot spots for which we have the condition $2\delta \ll R$ we can use the approximations

$$a = \sqrt{D^2 - R^2} = \sqrt{2\delta R} \; ; \; \sinh\eta_0 = \frac{a}{R} = \sqrt{\frac{2\delta}{R}} = \eta_0 \qquad (43)$$

We find that the critical parameter $\eta_0$ for the symmetric spherical dimers is $\eta_0 = \sqrt{\frac{2\delta}{R}} = \sqrt{\frac{d}{R}}$

where $d = 2\delta$ is the shortest distance between the surfaces of the two spheres. Although we treated special case we estimate that for more general dimers similar conclusions can be made.

## 6. Summary Discussion and Conclusions

In the present work we treated the mechanism by which "hot spots" are produced in the system of two metallic spheres with the same radius $R$ interacting with incident homogeneous EM field polarized in the symmetric $z$ direction. Hot spots with huge EM field are produced by local plasmons at a small gap with nanoscale dimensions. Such hot spots are measured by surface enhanced Raman spectroscopy (SERS) and two-photon induced luminescence (TPI-PL). These effects depend on the fourth power of the EM field at the hot spot where the measured molecules are inserted. While usually these effects are treated numerically or empirically the present theoretical analysis is based on the solution of Laplace equations by using bi-sherical coordinates with certain values for the dielectric constants of noble metals. In the present system the fourth power of the EM fields at the hot spot turns to have extremely large values when the shortest distance between the spheres surfaces $2\delta$ is very small i.e., when $2\delta \ll R$. Although we treated a very special system one can learn from such solutions on the general mechanism of hot spots.

Laplace equations with bi-spherical coordinates were developed in previous works [23-27] for obtaining the potentials at certain systems. The solutions turned to be very complicated involving many recursion relations with convergence problems. We developed in the present article certain approximations suitable for hot spots. In the present system in which the external EM is in the symmetric $z$ axis the potential has cylindrical symmetry about the $z$ axis. Therefore the potential $\psi_1(\eta,\alpha)$ at the hot spot developed in Eq. (10) is function of the bi-spherical coordinates $\eta, \alpha$, where $\eta$ represents the distance from the bi-spherical coordinates center and $\alpha$ represents an angle from this reference direction. The coordinates $\eta, \alpha$ can therefore be described as bi-spherical polar coordinates in the $x, z$ plane of Fig. 1, and these coordinates are not changed by rotation around the $z$ axis. The potential $\psi_1(\eta,\alpha)$ is proportional to summation of Legendre polynomials $P_n(\alpha)$ with proportionality coefficients $A_n$ and sinh function. The last term on the right side of Eq, (10) represents the external potential $V_{ext} = -Ez$ where $z$ is defined in bi-spherical coordinates in Eq. (2), and $E_0$ denotes, in short notation, the external EM field.

By using boundary conditions we obtained after some calculations and certain approximations (including the condition $\delta \ll R$) a general equation for the coefficients $A_n$ in Eq. (22). General solution for the EM field in the bi-spherical radial direction $\eta$ is derived in Eq. (26). Amplified EM field is found to be proportional to sum of products of the coefficients $A_n$ .with Legendre polynomial $P_n(\cos\alpha)$ and with cosh function. As the hot spots in dimers are produced on (or near) the symmetric $z$ axis, for which $x = y = 0$ we simplified the calculations by using this condition and used the relation: $\cos\alpha = -1$. The use of bi-spherical coordinates is demonstrated in Fig. 2.

In Section 5 we developed analytical results for the field enhancement factor ($EF$) at the center of the hot spot. Although the electric field has a complicated dependence on the coordinate $z$ for simplicity of calculation we used Eq. (34) for the hot spot center. The final results are given in Eq. (42) in which the light amplification factor is proportional to $\eta_0^{-5}$ and in Eq. (43) in which the EF for SERS measurements is proportional to $\eta_0^{-10}$ where $\eta_0 = \sqrt{\frac{2\delta}{R}} = \sqrt{\frac{d}{R}}$ and $d = 2\delta$ is the shortest distance between the two spheres. We estimate that for more general dimers the critical parameter will be the ratio of the gap length to the radius of curvature of the metallic surface near the gap

The present article is based on classical model but when the gap length is of an atomic scale quantum effects become important. Such quantum effects, including tunneling between the nanoparticles which reduces the *EF* , were treated, for example, in [39].